# UTFix: Change Aware Unit Test Repairing using LLM

SHANTO RAHMAN*, University of Texas at Austin, USA

SACHIT KUHAR, AWS, USA

BERK CIRISCI, AWS, Germany

PRANAV GARG, AWS, USA

SHIQI WANG, Meta, USA

XIAOFEI MA, ANOOP DEORAS, and BAISHAKHI RAY, AWS, USA

Software updates, including bug repair and feature additions, are frequent in modern applications but they often leave test suites outdated, resulting in undetected bugs and increased chances of system failures. A recent study by Meta revealed that 14%-22% of software failures stem from outdated tests that fail to reflect changes in the codebase. This highlights the need to keep tests in sync with code changes to ensure software reliability.

In this paper, we present UTFix, a novel approach for repairing unit tests when their corresponding focal methods undergo changes. UTFix addresses two critical issues: assertion failure and reduced code coverage caused by changes in the focal method. Our approach leverages language models to repair unit tests by providing contextual information such as static code slices, dynamic code slices, and failure messages. We evaluate UTFix on our generated synthetic benchmarks (Tool-Bench), and real-world benchmarks. Tool-Bench includes diverse changes from popular open-source Python GitHub projects, where UTFix successfully repaired 89.2% of assertion failures and achieved 100% code coverage for 96 tests out of 369 tests. On the real-world benchmarks, UTFix repairs 60% of assertion failures while achieving 100% code coverage for 19 out of 30 unit tests. To the best of our knowledge, this is the first comprehensive study focused on unit test in evolving Python projects. Our contributions include the development of UTFix, the creation of Tool-Bench and real-world benchmarks, and the demonstration of the effectiveness of LLM-based methods in addressing unit test failures due to software evolution.

CCS Concepts: • **Software and its engineering**; • **Computing methodologies** → *Software evolution, Artificial intelligence*;

Additional Key Words and Phrases: Software Testing, Unit Tests, Change Aware Test Repair, Large Language Models

---

*Work done while the author was an intern at AWS

---

Authors' Contact Information: Shanto Rahman, shanto.rahman@utexas.edu, University of Texas at Austin, USA; Sachit Kuhar, skuhar@amazon.com, AWS, USA; Berk Cirisci, cirisci@amazon.de, AWS, Germany; Pranav Garg, prangarg@amazon.com, AWS, USA; Shiqi Wang, tcwangshiqi@meta.com, Meta, USA; Xiaofei Ma, xiaofeim@amazon.com; Anoop Deoras, adeoras@amazon.com; Baishakhi Ray, rabaisha@amazon.com, AWS, USA.

---

## 1 Introduction

Software updates are common in most applications, usually to repair bugs or add new features. According to a recent Meta study, their central repository gets over 100,000 commits each week [7]. When these updates happen, developers should update their test code to check if the new features work as expected. But problems arise when the test code is left behind. Developers often neglect to keep their tests up to date, failing to evaluate changed scenarios and possible issues. This can lead to undetected bugs in production, increasing the chances of system failures that could have been avoided with better testing and maintenance. Meta's study shows that 14%-22% of failures come from outdated tests that do not reflect changes in the codebase [35]. These findings underscore the need of keeping tests in sync with code changes to ensure the reliability of software. In this paper, we focus on repairing unit tests when their corresponding focal methods are changed.

Software testing and reliability literature identify assertion failure and insufficient code coverage as key factors affecting reliability [8, 11, 20, 36]. Repairing these issues in evolving software is crucial, as ignoring or deleting failing tests compromises the test suite's effectiveness [11]. Moreover, a recent study by Microsoft found that tests are the most important area where developers need help from AI tools [20]. These findings highlight the need to continually update tests in evolving software to ensure the system remains robust and reliable.

When developers change a method, commonly referred to as the focal method (FM), the associated unit test can be affected in two primary ways: (i) existing test may fail, leading to an assertion failure, and (ii) parts of the changed FM may no longer be covered by the existing test. Repairing these tests is often time-consuming, and developers may not always prioritize it [11]. While significant work has been done on test generation [9, 17, 31–33], the area of test repair in evolving systems has received far less attention. The few studies that do address test repair tend to focus on specific sub-problems. For instance, ReAssert [11] concentrated solely on repairing test oracles, while Mirzaaghaei et al. [25] limited their repairs to specific changes, such as adding parameters or modifying variables and values. A more recent approach by Yaraghi et al. [42] applied a learning-based method to repair overall assertion failure but did not address improving code coverage of the changed FM.

To address the challenge of maintaining unit tests in an evolving codebase, we introduce UTFix, a novel approach designed to repair unit tests that have been affected by changes in the codebase. UTFix tackles both assertion failure and reduced code coverage caused by the changes in the FMs. When an FM changes, we begin by executing the existing, unmodified unit test to observe potential issues. This initial test run helps classify the nature of the failure: whether it is an assertion failure or a reduction in code coverage. If an assertion failure is detected—where the expected outcomes of unit test no longer match with the actual outcomes—UTFix attempts to pinpoint the root cause of the failure and repair it accordingly. For example, if the unit test results in reduced code coverage due to changes in the FM (i.e., certain branches or lines in the change code are not exercised), UTFix targets the uncovered lines or branches in the changed FM to repair the test.

Our primary contribution is UTFix, a systematic approach for repairing unit tests during code evolution by leveraging change impact analysis [5, 16]. UTFix modifies or adds assertions, updates test prefixes, and dynamically adapts tests to reflect changes in FMs. UTFix leverages Large Language Models (LLMs), providing them with essential contextual information such as static slices (code snippets relevant to the failure), dynamic slices (execution traces that may help to debug), and failure messages (details of what went wrong during the test run). With this information, we aim to enable LLMs to better understand the underlying problem and generate meaningful and syntactically correct test repairs. We evaluate UTFix in two ways:

(1) Synthetic Benchmark: To thoroughly understand different kinds of issues that might appear, we first curated a synthetic benchmark (Tool-Bench) with a diverse kind of changes. This Tool-Bench consists of 352 unit tests with assertion failures and 369 unit tests with reduced code coverage due to changes in the FMs from 44 projects. We found that UTFɪx can repair up to 314 assertion failures, achieving a repair rate of 89.2%. Additionally, 96 tests have 100% code coverage with an average coverage of 76.92%.

(2) Real-world Benchmark: To understand whether UTFɪx can address real-world changes, we further curated 50 real-world examples from eight heavily used GitHub projects, including 20 examples of assertion failures and 30 examples of reduced code coverage. We found that UTFɪx can repair 12 assertion failures, while 19 tests have 100% code coverage.

To the best of our knowledge, this is the first comprehensive study of test failures in an evolving software environment, with a particular focus on Python projects. While prior work has explored code failures and repairs, there is limited understanding of how continuous code evolution affects unit tests. Our study provides valuable insights into these test failures and introduces LLM-based methods to address them. Additionally, our approach incorporates novel prompting and feedback strategies, leveraging both static and dynamic slices to guide LLMs in effectively repairing unit test. The main contributions of our paper are:

- We propose UTFɪx to automatically repair unit tests that suffer from assertion failure and low code coverage due to changes in the corresponding focal methods.
- We create Tool-Bench and a real-world benchmark dataset for change aware unit test repair that have assertion failures and reduced code coverage.
- We implement UTFɪx for Python projects and evaluate it on both our Tool-Bench and the real-world benchmark dataset.

## 2 Motivating Example

In this section, we present two examples of how UTFɪx repairs unit tests: (1) when changes to the focal method cause an assertion failure, and (2) when changes to the focal method reduce code coverage.

### 2.1 Assertion Failure

```python
1 def bases(self, * ,force: bool = False):
2 -   return [self._base_from_info(
3 -     info) for info in self.
4 -     _base_info(force=force).bases]
5 +   base_info = self._base_info(
6 +     force=force).bases
7 +   if not force:
8 +     return [self._base_from_info
9 +       (info) for info in base_info]
10 +   else:
11 +     return [self._base_from_info
12 +       (info) for info in base_info if
13 +       info.id != 'appSW9R5uCNmRmfl6']
```

(a) Changed focal method

```python
1 def test_bases(...):
2   base_ids = [base.id for base in
3     api.bases()]
3   assert base_ids == ['appLkND', '
4     appSW9R5uCNmRmfl6']
4   ...
5   reloaded = api.bases(force=True)
6   assert [base.id for base in api.
7     bases()] == base_ids
7 - assert [base.id for base
8 -   in reloaded] == base_ids
9 + assert reloaded[0].id=='appLkND'
```

(b) Repaired test

Fig. 1. Assertion failure example

Figure 1 shows an example of assertion failure of unit test "test_bases" from the "gtalarico/airtable-python-wrapper" project. The left side shows the changed focal method (see Figure 1a), while the right side shows the corresponding repaired test (see Figure 1b). The changed focal method returns values based on the "force" parameter, excluding the ID 'appSW9R5uCNmRmfl6' when "force=True" (Lines 11-13 in Figure 1a). In the original test, the focal method is first called without specifying "force" (defaulting to "False") to retrieve all "base_ids" (Line 2 in Figure 1b). Later, the method is invoked with "force=True" at Line 5, outputting "reloaded" that excludes the specified ID. Retaining the original assertion (Lines 7-8) then leads to an assertion failure since "reloaded" no longer matches "base_ids".

However, the repaired test addresses this issue by adjusting the assertion logic at Line 9 in Figure 1b. Rather than comparing all base IDs, the updated test explicitly verifies if the "reloaded" ID is equal to 'appLkND', aligning with the changed focal method's logic to filter out certain base IDs. This prevents the assertion failure and ensures the test accurately reflects the changed behavior of the focal method.

## 2.2 Reduced Code Coverage

Figure 2 shows another example from the same project ("gtalarico/airtable-python-wrapper") used in our evaluation where code coverage is reduced due to the changes in focal method. Figure 2a shows an example of changed focal method, while Figure 2b shows the corresponding repaired test.

Figure 2a shows that the focal method "batch_delete" introduces a check for empty "record_ids" during its evolution. Previously, it processed IDs in chunks via "self.api.chunked(record_ids)" at Line 6, sending DELETE requests to "self.url" for each chunk, validating results with "assert_typed_dicts" and aggregating them into "deleted_records" at Line 8. The changes in the focal method adds a condition at Line 4 to immediately return an empty list if "record_ids" is empty, thereby avoiding unnecessary API calls in subsequent lines of code.

```python
1  def batch_delete(...):
2    deleted_records = []
3    record_ids = list(record_ids)
4  + if not record_ids:
5  +   return deleted_records
6    for chunk in self.api.chunked
       (record_ids):
7      result=self.api.delete(self
         .url, params = {'records
         []': chunk})
8      deleted_records +=
         assert_typed_dicts(
         RecordDeletedDict, result
         ["records"])
9    return deleted_records
```

(a) Changed focal method

```python
1  def test_batch_delete(...):
2    ids = [i['id'] for i in mock_records]
3    with Mocker() as mock:
4      for chunk in _chunk(ids, 10):
5        json_response= {'records':[{'deleted'
           : True, 'id' : j} for j in chunk]}
6        url_match = Request('get', table.url)
7        mock.delete(url_match, status_code =
           200, json = json_response)
8  +     empty_resp = table.batch_delete([])
9  +     assert empty_resp == []
10       resp = table.batch_delete(ids)
11     expected = [{"deleted": True, "id": i}
         for i in ids]
12   assert resp == expected
```

(b) Repaired test

Fig. 2. Code coverage example

Figure 2b shows that "test_batch_delete" has been repaired to reflect changes in the focal method and ensure full code coverage. A new assertion at Line 9 verifies that "batch_delete([])" returns an empty list, while the original assertion (Line 12) still validates non-empty inputs. This ensures 100% code coverage by covering all execution paths.

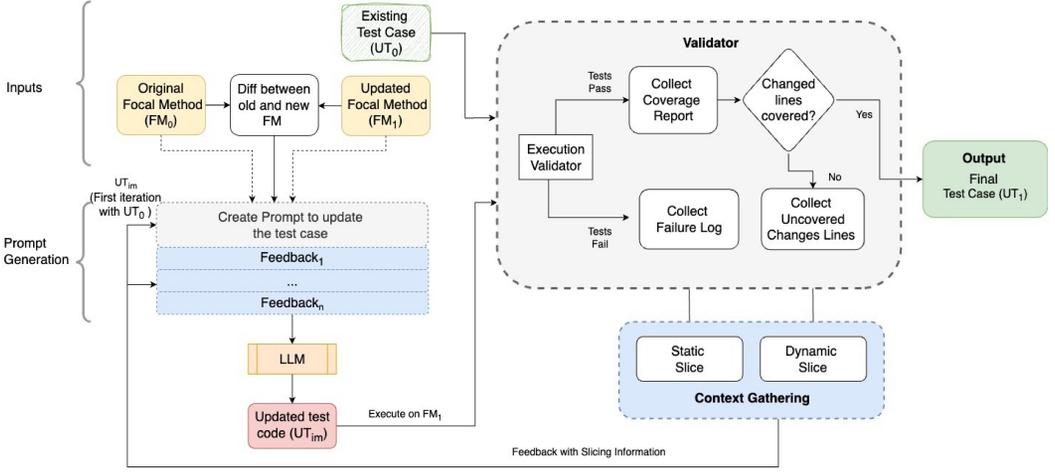

Fig. 3. UTFix Framework

## 3 Methodology

In this section, we introduce UTFix, an approach to repairing unit test for a focal method that has undergone changes. Test repair addresses two main scenarios: (i) assertion failure, where an existing unit test fails due to an assertion check, and (ii) reduced code coverage, where the existing test passes but covers less code than before. At the core of our approach is a multi-stage prompting with feedback that breaks the problem of test repair into step-by-step debugging and repairing steps. Similar to how a developer might debug a test failure, we create the prompt based on failure logs, the static slices of the focal method, and the paths that the test has traversed (using dynamic slices). These prompts are designed to instruct the model to repair the existing test based on the observed failures. Figure 3 shows an overview of UTFix.

### 3.1 Terminologies

First, we introduce some key terminologies that we will follow throughout the paper.

*Focal Method (FM).* This is the method under test. In our setting, the focal method has been updated from version 0 ($FM_0$) to version 1 ($FM_1$). The changes between the two versions can be represented by their diff $\Delta_{(FM_0, FM_1)}$.

*Unit Test (UT).* An FM is associated with a unit test, $UT$, to evaluate its implementation. We call the unit test associated with $FM_0$ as $UT_0$, and when the focal method is updated to version 1, the corresponding updated test is termed as $UT_1$.

*Static Slicing.* The traditional definition of program slicing [38] of a variable involves identifying all statements in a program that directly or indirectly affect the value of the variable. In the context of focal method, we adapt this definition as follows: static slicing refers to all the statements within the focal file (where focal method resides) that can directly or indirectly impact the function's call and return values.

*Dynamic Slicing.* The above concept of static slices considers all potential statements that may impact focal method but does not account for the specific inputs (in our case, tests) that cause the failure. However, developers typically examine a failure log in relation to the test that triggered the failure. Agrawal et al. [6] introduced dynamic slicing to analyze program behavior with respect to specific bug inducing inputs. Building on this, in this paper, we define the dynamic slices of a focal method with respect to a unit test as all the statements within the focal method and in the methods directly called by the focal method during the test's execution.

## 3.2 Overview

Figure 3 shows a high-level framework of UTFıx. It starts by taking original and changed focal methods, $FM_0$ and $FM_1$, respectively, and the original test ($UT_0$) as input. Next, $UT_0$ goes through a *validation* step— if $UT_0$ can execute $FM_1$ successfully without any assertion failure and demonstrates no regression in code coverage compared to $FM_0$, no further action is needed, and UTFıx will stop at this point. However, if $FM_1$ results in an assertion failure or shows a reduction in code coverage, UTFıx initiates the repair process.

For assertion failure, UTFıx gathers the failure log, which captures the details of the test execution and failed assertions. For coverage reduction, UTFıx collects the uncovered lines in $FM_1$. In addition, UTFıx collects both static and dynamic slices for $FM_1$ during its execution with the test. Static slices capture code segments (methods/statements) within the focal file that could potentially impact $FM_1$'s call and return values, while dynamic slices focus on statements that were actually executed by $UT_0$ and impacted the behavior of $FM_1$. Static slices are obtained by prompting a Large Language Model (LLM) with relevant context, such as file structure and method definitions. Dynamic slices, on the other hand, are derived by analyzing the coverage report, which provides execution paths and statement coverage. The collected slices, along with failure logs and changes in the focal methods, are then used by another LLM to pinpoint which statements and methods are responsible for the observed failures, enabling a focused and informed repair process.

UTFıx then consolidates this information and prompts the LLM to repair the failed test. It provides the LLM with additional context, including the original ($FM_0$) and the changed focal method ($FM_1$), highlighting their differences. Once the LLM generates an updated test (a.k.a., intermediate test $UT_{im}$), UTFıx executes $UT_{im}$ against $FM_1$ and performs a similar *validation* process. If $UT_{im}$ passes the validation, it becomes the final repaired test. If not, the feedback and repair loop is repeated, continuing up to a predefined threshold times of repetition.

To this end, our test repair contains three steps: (Step-I) validating a test when the corresponding focal method has evolved and collect feedback, (Step-II) capturing relevant code slice, and (Step-III) generating prompts for an LLM by leveraging feedback and context information collected in the above step.

## 3.3 Step-I: Test Validation and Feedback Collection

*3.3.1 Validator.* Our approach begins with validating the given test, $UT_0$, against the changed version of the focal method, $FM_1$. The goal of this initial validation step is to determine whether the test fails when executed with the $FM_1$. The validator runs $UT_0$ with $FM_1$, checking for both assertion failures and any potential regressions in code coverage as compared to running $UT_0$ with $FM_0$. If the test passes without any issues, indicating that it successfully covers the changes in $FM_1$, we consider the test as updated and output it as the final version, concluding the process. Otherwise, our approach proceeds to collect feedback.

*3.3.2 Feedback.* When an assertion failure occurs, we collect the failure log generated during the test execution. This log contains essential information about what went wrong, including which assertions failed and the conditions under which it failed. We then parse the log to extract the most relevant sections, focusing on the specific parts of the code and test execution that led to the assertion failure. By isolating these key details, we can better understand the root cause of the issue and prepare for the next step: gathering context and generating prompts to repair the test.

```
def test_bases(api, mock_bases_endpoint):
    base_ids = [base.id for base in api.bases()]
    assert mock_bases_endpoint.call_count == 1
    assert base_ids == ["appLkNDICXNqxSDhG", "appSW9R5uCNmRmfl6"]

    # Should not make a second API call...
    assert [base.id for base in api.bases()] == base_ids
    assert mock_bases_endpoint.call_count == 1
    # ....unless we force it:
    reloaded = api.bases(force=True)
>   assert [base.id for base in reloaded] == base_ids
E   AssertionError: assert ['appLkNDICXNqxSDhG'] == ['appLkNDICXN...9R5uCNmRmfl6']
E
E     Right contains one more item: 'appSW9R5uCNmRmfl6'
E
E     Full diff:
E       [
E           'appLkNDICXNqxSDhG',
E       -   'appSW9R5uCNmRmfl6',
E       ]
```

Fig. 4. Example of an assertion failure, highlighting the lines taken as feedback. Only the failure points starting with '>' and lines starting with 'E' as 'Error' tag are considered to pinpoint the issue.

This enables more targeted feedback for the LLM, helping it make more precise adjustments to the test. Figure 4 shows an example of a failure log.

If $UT_0$ shows reduced code coverage while running on $FM_1$, we mark the lines in $FM_1$ that are not covered by $UT_0$. When $FM_0$ is updated to $FM_1$ by adding new branches or method calls, certain statements in $FM_1$ may not be covered by the original test $UT_0$. Furthermore, if a new branch condition is added, previously covered lines may no longer be covered by $UT_0$. We mark all these lines and send them as feedback to test repair phase. To ensure better code coverage, it is sometimes necessary to generate new tests targeting these uncovered lines instead of updating the existing ones.

### 3.4 Step-II: Collection of Code Slices

UTFix collects static and dynamic slices of the changed focal method ($FM_1$) during its execution with the test. These slices provide essential insights into the code that might have contributed to the test failure. For instance, static slices include statements within the focal file and its associated methods that could potentially impact $FM_1$'s call and return values. This means that, regardless of whether these statements were executed during the test run, they could directly or indirectly influence the behavior of $FM_1$ in future executions. The static slice provides a broader picture of the potential dependencies in the code. Dynamic slices, on the other hand, focus only on the statements that were actually executed during the test run. These include the specific lines of code that $UT_0$ triggered while executing $FM_1$. By narrowing down to the exact execution paths that were taken during the test, the dynamic slice gives a precise view of what may have directly influenced $FM_1$'s behavior and caused the failure.

To gather static slices, UTFix prompts an LLM, providing it with a detailed context. The LLM uses this context to identify the relevant code that either might or actually did contribute to the failure. We provide the $FM_1$ and focal file code, and ask LLM to generate the most relevant code slices from this focal file. The LLM produces one or more relevant code slices related to $FM_1$. Figure 5 shows an example output of static slice that language model generates. We find that the static slice primarily identifies related methods or classes of the focal method within the focal file. For instance, as shown in Figure 5, the UTFix's static slice analyzer returns two relevant contexts. In this case, the focal method is "bases", same as the example shown in Figure 1. The analyzer identifies the methods such as "bases", and "_base_from_info", as shown in <context-1>, and <context-2>, respectively, that are called from the focal method.

```
1  <context-1>
2      def _base_info(self) -> Bases:
3          """ Return a schema object that represents all bases."""
4          url = self.build_url("meta/bases")
5          data = {"bases": [
6                  base_info
7                  for page in self.iterate_requests("GET", url)
8                  for base_info in page["bases"]
9                  ]
10          }
11          return Bases.from_api(data, self)
12  </context-1>
13  <context-2>
14      def _base_from_info(self, base_info: Bases.Info):
15          return pyairtable.api.base.Base(self, base_info.id, name =
            base_info.name, permission_level = base_info.permission_level)
16  </context-2>
```

Fig. 5. An example of static slice

We further collect the dynamic slices by running $UT_0$ on the focal method and saving the code coverage. To collect code coverage, we first automatically instrument the "tox.ini" file (tox is an automated tool primarily used in Python to run tests) using pytest-cov plugin. This plugin keeps track of which lines are executed during a test run. Then we run our intended test in isolation through tox that generates a coverage report. This report only contains the line numbers of the Python files that are covered but does not indicate relationships between lines and methods. To address this, we traverse the AST of each method of all the covered Python files and intersect them with the coverage report to compute the coverage per method. However, this often contains excessive information unrelated to the changes in the focal method. So, we minimize the dynamic trace by focusing on the focal method's executions and the methods it calls. This results in a reduced dynamic trace, which we refer to as dynamic slices. Importantly, we minimize the information provided to avoid overwhelming the model, as too much data can lead to overestimation and the generation of unrelated test code.

We find that the dynamic slice often offers valuable insights into code execution. For example, in Figure 6, we observe that all the methods called from the focal method are captured, but only the lines executed during the test run are extracted. When compared to static slice retrieval, the "_base_info" method includes many additional lines. Our experiments show that providing excessive, unrelated information from the static slice can often mislead the performance of language model.

By pinpointing these critical static and dynamic slices, the language model is better equipped to assist in generating an effective repair. This focused and informed approach allows the repair process to target the root causes of the issue, improving the chances of successfully repairing the test without unnecessary changes.

### 3.5 Step III: Prompt Generation with Context and Feedback

After gathering the failure log and the static and dynamic slices, UTFɪx combines all this information and uses it to prompt the LLM for unit test repair. In this step, UTFɪx provides the LLM with a comprehensive context to aid in generating an effective repair. This context includes not only the failure details but also both the original focal method ($FM_0$) and the changed focal method ($FM_1$), along with their differences (the diff, as shown in Figure 1). By supplying the LLM with

```
1  <filename name="pyairtable/api/api.py">
2    <method name="_base_info">
3      <method_body>
4        def _base_info(self) -> Bases:
5          url = self.build_url("meta/bases")
6          data = {
7            "bases": [
8              ...
9              ]
10           }
11         return Bases.from_api(data, self)
12     </method_body>
13   </method>
14   ...
15 </filename>
16 <filename name="pyairtable/api/base.py">
17   <method name="__init__">
18     <method_body>
19       def __init__(
20         self.api = api
21         self.id = base_id
22         ...
23     </method_body>
24   </method>
25 </filename>
```

Fig. 6. An example of dynamic slice. This includes all the methods that are called from the focal method.

these elements, it helps the model better understand the underlying changes that led to the failure, allowing it to focus on how the test needs to be adjusted.

The LLM, based on this context, generates an updated test ($UT_{im}$) that is designed to pass the changed focal method ($FM_1$). Once the updated test is generated, UTFix verifies that the output is in the correct format, ensuring that all opening and closing tags are present and complete. It also checks for syntax errors in the generated test, such as unmatched opening and closing brackets. UTFix then automatically runs the test against $FM_1$, evaluating its correctness and code coverage. If $UT_{im}$ passes the test without any assertion failure and meets the desired coverage requirements, it is accepted as the final repaired test, and the process concludes.

However, if $UT_{im}$ still fails or does not meet coverage expectations, the process enters a feedback loop. In this loop, UTFix refines its inputs by providing the LLM with the results of the failed attempt, along with additional failure logs and updated dynamic slices if necessary. The LLM is then prompted again to adjust or further refine the test based on the new information.

This feedback-and-repair loop continues iteratively, allowing the LLM to progressively improve the test. The process repeats until a successful test is generated or a predefined threshold of attempts is reached. If the threshold is reached without success, the process stops, signaling that further intervention may be needed.

## 4 Benchmark Creation

In this section, we develop a synthetic benchmark dataset (Tool-Bench), and a real-world benchmark dataset for evaluating UTFix. First, we collect unit tests and their corresponding focal methods from popular GitHub repositories. For Tool-Bench, we introduce realistic changes into the focal methods and assess UTFix's ability to accommodate those changes in the corresponding unit tests. Besides the synthetic benchmark dataset, we create a real-world benchmark dataset from open-source GitHub projects that searches <FM, UT> change pairs in the commit histories allowing us to evaluate UTFıx's ability to accurately repair tests in real-world software evolution.

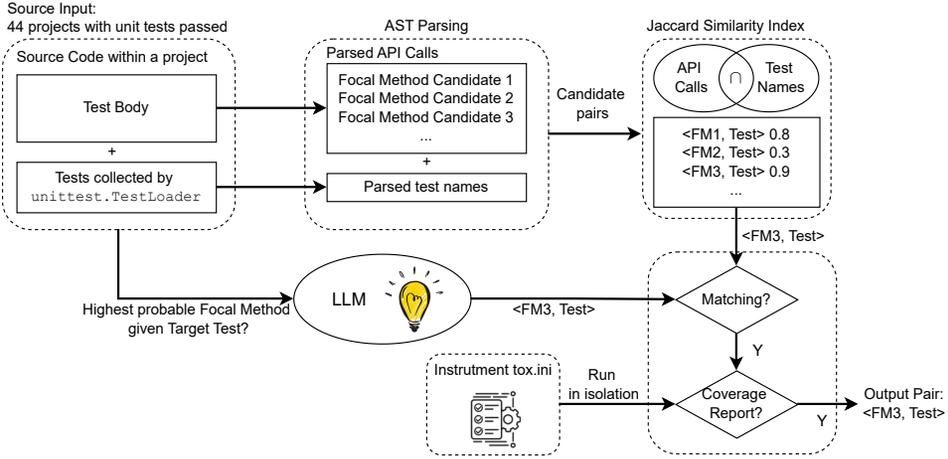

Fig. 7. Tool-Bench: a synthetic benchmark designed to identify <FM, UT> pairs. To identify the primary focal method (FM) for each targeted test, we leverage two techniques: (1) Jaccard Similarity Index with AST parsing and (2) predictions from an LLM. For quality assurance, we only consider pairs where both approaches produce consistent results, yielding 1072 <FM, UT> pairs. Finally, the tests are run in isolation, and code coverage reports are generated, resulting in 690 <FM, UT> pairs.

## 4.1 Tool-Bench

In this synthetic data curation process, we have four major steps: i) collecting unit tests from open-source GitHub projects, ii) identifying <FM, UT> pairs, iii) validating <FM, UT> pairs, and iv) introducing changes to the FM to evaluate the performance of UTFıx. The details of each step are described as follows.

*Step 1: Collecting Unit Tests from GitHub projects.* We started by collecting popular open-source Python projects from GitHub. Next, we ran each project and filtered out the projects that have compilation or build errors, resulting in 44 projects, as illustrated in Figure 7. These projects span a variety of categories, including Programming and Code Management, Data Processing, Django-related development, Web and API Development, Data Parsing, and other miscellaneous topics. Finally, we collected unit tests from these projects using the "unittest.TestLoader" framework.

*Step 2: Identifying <FM, UT> Pairs.* A unit test often invokes multiple methods for setup or post-processing, in addition to the primary focal method call. For example, in Figure 2, although the test "test_batch_delete" is primarily testing the focal method "batch_delete", it also calls other methods like "delete". The presence of this multiple method invocation increases the complexity of finding the primary focal method. To address these complexities, we employ two distinct techniques such as name similarity analysis and leveraging an LLM. We then identify the primary focal methods as those recommended by both techniques, as illustrated in Figure 7.

- Leveraging Name Similarity Analysis: After collecting the unit tests, we aim to identify a primary focal method that each test  is running. We analyze the Abstract Syntax Tree of each test method to extract all method calls that it invokes. For each unit test, we denote the extracted method calls as $\{FM_1, FM_2, \ldots, FM_k\}$, and apply Jaccard Similarity Index to identify the most similar method call to the test name. Hence, we first tokenize both the test name and all method calls that occur within the test. Standard text processing techniques, such as splitting camel case identifiers and handling underscores or dashes, are applied to generate token lists for each focal method. Let $T_L$ represent the token list for the test name, and $FM_{iL}$ represent the token list for each method call. The Jaccard Similarity Index between the test name and each method call is computed as:

$$FM_J = \arg\max_{i \in k} \frac{|T_L \cap FM_{iL}|}{|T_L \cup FM_{iL}|} \tag{1}$$

Equation 1 illustrates how we select the focal method based on the token list of test method name ($T_L$) and the token list of each method calls ($FM_L$). The method call $FM_i$ with the highest similarity score to the test method is considered the primary focal method ($FM_J$) for the given test method.

- Leveraging LLM: We leverage an LLM to further enhance the identification of valid <FM, UT> pairs. Given a test name and its corresponding test body, the language model is prompted to provide the primary method calls, $\{FM_1, FM_2, \ldots, FM_k\}$, along with their probability scores, $P(FM_i)$. We then select the method call $FM_{LLM}$ as the primary focal method with the highest probability score, where:

$$FM_{LLM} = \arg\max_{FM_i} P(FM_i)$$

After obtaining the focal method from the two techniques mentioned above for a given test, we identify it as the primary focal method if both techniques suggest the same focal method, as defined in Equation 2. Otherwise, we consider that the focal method identification has failed for that test. This process results in 1072 <FM, UT> pairs.

$$FM = \{FM_J\} \cap \{FM_{LLM}\}, \ FM \neq \emptyset \tag{2}$$

*Step 3: Validating <FM, UT> pairs.* In this step, we run each test to confirm that the considered FM is actually called from the UT. Hence, we modify "tox.ini" file that is associated with the tox automation tool primarily used in Python projects to run tests in different Python versions. Then we run each test in isolation. We collect code coverage and trace all methods under test, denoted as $FM_{CC} = \{fm_1, fm_2, \ldots, fm_k\}$, that were executed during the test run. We identify a valid focal method by matching each method from code coverage $FM_{CC}$ with the previously obtained candidate focal method $FM$ from *Step 2*. A method $FM$ is considered as valid if:

$$FM \in FM_{CC}$$

This condition ensures that the focal method ($FM$) is invoked during the test run. Finally, we obtain 690 valid <FM, UT> pairs. Other sophisticated approaches, such as static slicing or data flow analysis, could be used to identify <FM, UT> pairs within the selected packages. However, as demonstrated through our experiments in Section 6, our current approach for constructing <FM, UT> pairs neither compromises the validity of our experimental results nor introduces bias into the dataset.

*Step 4: Curating synthetic changes in FM.* After identifying <FM, UT> valid pairs, we curate changes to the focal method (FM) of each pair to evaluate whether UTFix performs well on the change aware settings. To curate the changes, we leverage an LLM. The inputs to the LLM include the focal method body, the focal method file contents, the test file contents, and the unit test body. This context helps to guide the LLM in generating meaningful changes to the focal method. When making changes to focal method, our goal is to introduce a diverse set of changes. We instruct the LLM to apply 51 different types of changes. Each change is designed to impact the test outcome while maintaining code compilability. Then we automatically inject the newly curated $FM_1$ by replacing the existing focal method ($FM_0$). Following these steps, we generate two datasets: one for assertion failure caused by changes in the focal method, and another for low code coverage caused by changes in the focal method while ensuring the test still passes.

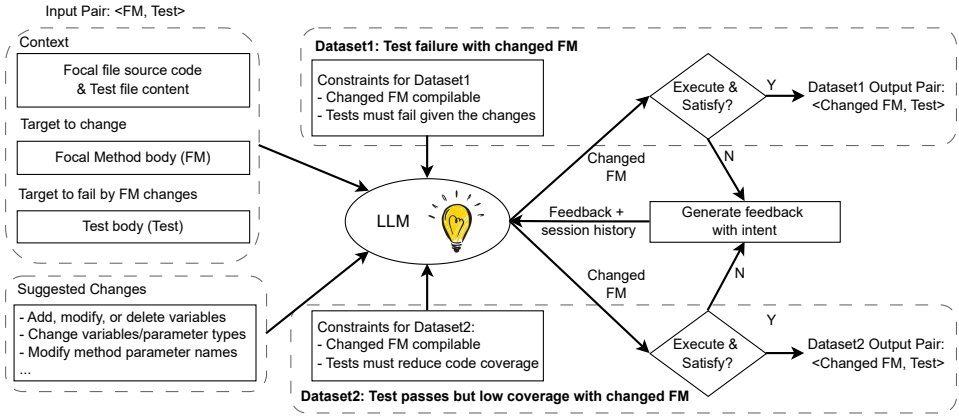

Fig. 8. Overview of the focal method change process using LLM to generate two benchmark datasets. The first dataset focuses on changing the focal method to trigger assertion failure in the corresponding test. The second dataset aims to change the focal method to reduce code coverage while ensuring that the test still passes. Both datasets are curated using a similar approach, with adjusted constraints to meet their respective objectives.

- Assertion failure due to focal method change: To generate the assertion failure dataset, we run the test and observe if the changes in focal method result in an assertion failure. If the curated method $FM_1$ satisfies this condition, we consider the change as a valid change in focal method. Otherwise, if changes to the focal method result in compilation errors, syntax errors, or module-not-found errors, we discard those changes and use chain-of-thought (COT) as a feedback loop. Figure 8 shows the workflow for generating change aware benchmark dataset. During the feedback loop, we generate the prompt by parsing the error log as input, and update the "Chat History" with new contextual information. We apply these iterative changes in FM for three times. However, Tool-Bench terminates the feedback loop sooner if it observes an identical compilation error in the consecutive rounds, resulting in 352 unit tests with assertion failure.
- Test passes but focal method change reduces code coverage: To generate the reduced code coverage benchmark dataset, we run the test with the changed focal method and check if it still passes but with reduced code coverage. This is important because insufficient code coverage poses a risk of test failure in a CI/CD environment. Each time a change occurs in the focal method, we run the corresponding unit test and calculate the code coverage for that focal method. If the code coverage is lower than it was before the focal method was changed, we consider the change as a valid one and include it in our dataset. Otherwise, we use COT as a feedback loop. As a result, we get 369 tests with reduced code coverage.

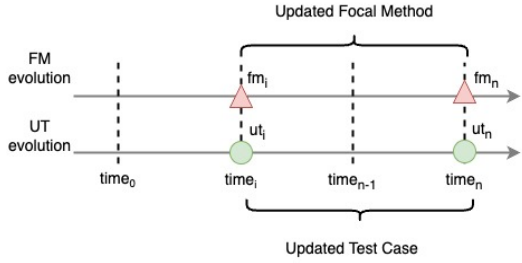

Fig. 9. Overview of data curation for real-world changes of focal method ($fm$) and corresponding unit test ($ut$). It shows their evolution in the project timeline from $time_0$ to $time_n$. In particular, $fm$ is updated from version $fm_i$ to $fm_n$, and the corresponding $ut$ ($ut_i$ to $ut_n$). We collect the diff between the two versions of $fm$ and $ut$. If both undergo changes, and the old version of $ut$ passes old version of $fm$, but fails new version of $fm$, we collect such change pairs in our dataset.

## 4.2 Curating real-world changes

For real-world data collection, we begin by identifying a test and a focal method pair from a specific commit (typically, the latest SHA). This provides a snapshot of the project's current state, including the interaction between the test and the focal method at the time of our study. We then walk backward through the project's evolution timeline to curate <FM, UT> changes. Figure 9 illustrates this process.

*Step 1: Commit history traversal.* We traverse the commit history of the project, moving backward in time. By collecting all previous commits on focal method ($FM$) and unit test ($UT$), we capture their complete historical context in the codebase. In particular, for each commit, we extract the pre- and post-change versions of the $FM$ and $UT$. This extraction allows us to observe how the <FM, UT> evolved between consecutive commits.

*Step 2: Detecting changes in focal method and test.* To detect changes, we compare the Abstract Syntax Trees (ASTs) of the pre- and post-change versions of the $FM$ and $UT$. Using AST comparisons, we detect structural changes in the code rather than relying on textual differences, which may not capture deeper syntactical changes. For example, refactorings that move code blocks or change method signatures are better reflected through AST changes. Next, we check whether any significant changes have occurred in either the focal method or test method. If no changes are detected, we discard the commit from further analysis. The rationale here is that unchanged commits do not provide meaningful data evaluating the effectiveness of UTFix. Otherwise, we mark the changes for further processing. For instance, in Figure 9, $fm_i$ and $fm_n$, and $ut_i$ and $ut_n$ show differences, and we mark those for further studies.

*Step 3: Validity of commit (no assertion failure).* We evaluate the validity of each commit by checking whether the corresponding test at that commit continues to pass. A commit is considered valid only if there is no assertion failure at that commit. For example, $ut_i$ should pass $fm_i$ for further consideration.

*Step 4: Replacing current test and evaluating failures.* After identifying a valid commit, we return to the latest commit and replace the test method in the current codebase with its body from the older commit (e.g., $ut_n$ will be replaced by $ut_i$ as shown in Figure 9). This allows us to simulate how the previous version of the test behaves when executed against the newer focal method (e.g., $fm_n$). If this replacement causes assertion failure or reduced code coverage, we include that case in our dataset. The rationale behind this step is to make sure that the old test could not adapt to the changes in the focal method and requires repair. We then evaluate UTFix's repair capability on the old test ($ut_i$).

*Step 5: Sampling for real-world changes.* From the collected data, we sample 20 and 30 real tests that result in assertion failures and reduced code coverage, respectively. Assertion failures indicate a direct violation of the expected behavior, whereas reduced code coverage indicates that certain parts of the code are no longer tested due to changes in the focal method. Both types of changes are critical for evaluating how well UTFix adapts tests to evolving code.

## 5 Experimental Setup

### 5.1 Research Questions

To evaluate the effectiveness of UTFix, we address the following research questions:

- RQ1: How effective is UTFix in repairing failed tests?
- RQ2: How effective is UTFix in generating tests to improve the code coverage of the focal method?
- RQ3: How much time does UTFix require to repair or generate tests?
- RQ4: What is the diversity of changes in our benchmark datasets and the repaired tests?
- RQ5: How does UTFix perform when applied to real-world code changes?

We address RQ1 to see whether the intuition behind UTFix repair strategies is effective for repairing unit tests when changes occur in focal methods. We address RQ2 to evaluate the effectiveness of UTFix in generating new tests to improve code coverage. We address RQ3 to show the cost of running UTFix. We address RQ4 to examine the characteristics of changes in our benchmark dataset and repaired tests. We address RQ5 to show how good UTFix is at repairing unit tests when real changes occur in focal methods.

### 5.2 Evaluation Dataset

We collected 44 popular open-source GitHub projects. From these projects, we curated benchmark datasets following the steps of Tool-Bench that resulted in 352 tests with assertion failures, and 369 tests with reduced code coverage as discussed in Section 4. Assessing the quality of AI-generated data remains an open challenge, though research highlights that careful prompting can significantly enhance data quality [22]. Building on this approach, we ensure quality across four dimensions:

- Syntactic and Semantic correctness: All curated changes are verified as compilable by running tests.
- Realistic Development Scenario: For benchmark dataset with reduced code coverage, we verify that tests pass while achieving less than 100% coverage, reflecting realistic scenarios of incomplete testing.
- Diversity: We prioritize varied focal method changes, including 'error handling', 'API updates', 'branches', 'loops', and 'data structure' changes.
- Manual Checking: Additionally, we conducted a manual spot check on a subset of the data to ensure its quality (e.g., obvious hallucination is not present, etc.).

We also conduct experiments to curate real-world changes from eight GitHub projects by traversing their commit histories, yielding 20 examples of assertion failures and 30 examples of reduced code coverage.

### 5.3 Running Environment

We run unit tests in a "tox" [4] environment with Python version 3.9. We use Claude 3.5 Sonnet model [1]. To implement chain-of-thought reasoning, we maintain "ChatMessageHistory" from "langchain community.chat message histories" [3]. We set the temperature to 0.9 for creating Tool-Bench. For repairing unit tests, we set the temperature to 0.4, the top_k as 0.4, and top_p as 250. For collecting coverage report, we use "pytest - -cov". We run our experiments in Ubuntu 20.04 with 4 CPUs, 8GiB RAM, and no GPU required for test repairs. On average, repairing each test costs approximately $0.0034.

## 5.4 Evaluation Settings

Let $FM_0$ and $UT_0$ correspond to old code versions, and $FM_1$ and $UT_1$ correspond to the new versions. We evaluate UTFix with the following ablation settings. 'error' indicates assertion failure message or uncovered lines.

- Baseline: $FM_1$ + error + $UT_0$
- UTFix with no slice (UTFix$_{nc}$) : $\Delta(FM_0, FM_1)$ + error + $UT_0$
- UTFix with static slice (UTFix$_s$) : UTFix$_{nc}$ + static slice w.r.t. $FM_1$
- UTFix with dynamic slice (UTFix$_d$) : UTFix$_{nc}$ + dynamic slice while running $UT_0$ on $FM_1$
- UTFix with both static and dynamic slice (UTFix$_{s+d}$): UTFix$_{nc}$ + static slice + dynamic slice

**Baseline Selection.** Since no existing tools directly address the problem in our specific setting, we implement a reasonable baseline approach that simulates how a developer might attempt to repair a failing test. This baseline involves crafting a detailed prompt for the LLM with all the necessary information to facilitate a repair.

To repair a test ($UT_0$), our baseline approach constructs a prompt that includes three key elements:

- The Updated Focal Method ($FM_1$): We include the full body of $FM_1$ to ensure that the LLM has the context of the changes that the test needs to accommodate.
- The Error Produced by $UT_0$: We include the exact error that occurred when running the test. This helps the LLM to understand the nature of the problem—whether it is an assertion failure or code coverage—and allows it to focus on the specific error.
- The Original Unit Test ($UT_0$): The baseline also provides the LLM with the body of the failing unit test. This gives the LLM the context of how the test is currently written, allowing it to update the test in a way that aligns with the new version of the focal method.

The idea behind this baseline is to give the LLM enough context to generate a reasonable repair for the failing tests. It mimics how a knowledgeable developer might use an LLM in a practical setting, where they provide the relevant portions of the code and the error message in order to receive a repair suggestion. We also evaluated an enhanced version of this baseline by incorporating multiple rounds of feedback, similar to our overall framework. This iterative feedback loop ensures that the LLM can continue improving the test until it passes or reaches an optimal state, further aligning this baseline with how users might practically employ LLMs for continuous test repair.

This baseline serves as a point of comparison for evaluating UTFix, helping us to assess how much value our approach adds beyond what a user might achieve through simple prompting and feedback with an LLM.

## 6 Results

### 6.1 RQ1: Effectiveness in Repairing Unit Tests Causing Assertion Failures

Figure 10 shows the number of repaired tests by UTFix compared to the Baseline. As UTFix has four different settings: UTFix with no slice (UTFix$_{nc}$), UTFix with static slices (UTFix$_s$), UTFix with dynamic slices (UTFix$_d$) and UTFix with both static and dynamic slices (UTFix$_{s+d}$), we show for each as four separate lines in the chart, along with one additional line for the Baseline. Figure 10 shows the effectiveness of using the diff of the changed focal method (UTFix) compared to using the full changed focal method (Baseline). The Y-axis of this figure (line chart) shows the cumulative number of repaired tests, and the X-axis shows five different feedback passes (chain-of-thought (COT) counts), including zero-shot (i.e., Pass@0), which are used to repair the tests. Since we are using COT five times, we report the number of tests repaired at each pass in Figure 10.

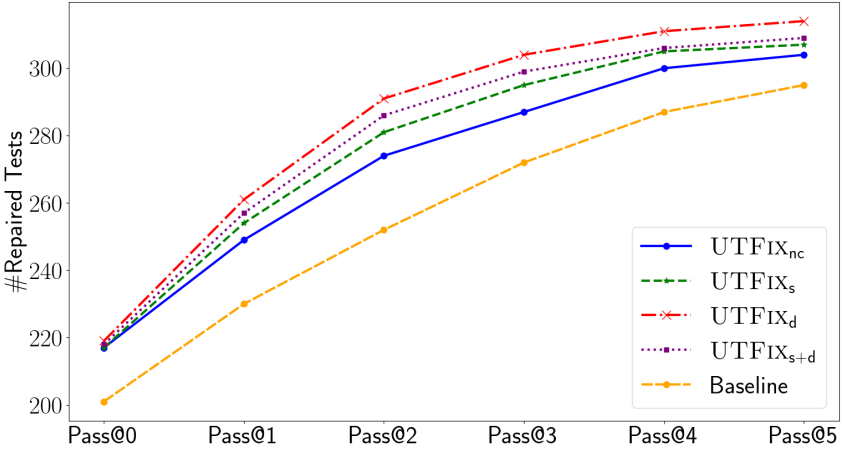

Fig. 10. Cumulative test repairs over five feedback iterations for all four different settings of UTFɪx, snd Baseline. The figure shows the UTFɪx always outperforms the Baseline. Also when considering dynamic slices, UTFɪx can repair maximum number of tests.

Figure 10 shows that all four UTFɪx techniques (with different colors) achieve nearly identical results at Pass@0, with 217, 217, 219, and 218 tests repaired. However, this number varies when static or dynamic slices are incorporated into the feedback. Specifically, the feedback with dynamic slices (marked in red) is the most effective, repairing 42 tests at Pass@1, compared to 39, 37, and 32 with UTFɪx$_{s+d}$, UTFɪx$_s$, and UTFɪx$_{nc}$, respectively. We find that UTFɪx with dynamic slices outperforms the other settings, repairing a total of 314 out of 352 test failures (89.2% tests). In comparison to the Baseline, UTFɪx consistently performs better, as the Baseline repairs a maximum of 295 test failures.

Interestingly, the combination of both static and dynamic slices does not consistently outperform dynamic slices alone. This is likely because the combined feedback contains excessive, irrelevant, and conflicting information, which hinders the model's ability to repair tests effectively. Static slices include all possible paths, while dynamic slices focus only on the paths that are actually executed. Their combination may sometimes introduce noise and irrelevant features, making it harder for the model to generate repaired tests as accurately as when using only the dynamic slices. Additionally, having these different types of slices sometimes complicates data interpretation, preventing the test repair process.

## 6.2 RQ2: Effectiveness in Improving Code Coverage When Code Changes Reduce Coverage

Figure 11 shows the code coverage achieved by six different techniques: $UT_0$, Baseline, UTFɪx with no slice (UTFɪx$_{nc}$), UTFɪx with static slices (UTFɪx$_s$), UTFɪx with dynamic slices (UTFɪx$_d$), and UTFɪx with both static and dynamic slices (UTFɪx$_{s+d}$). We show the results using different violin plots, as shown in Figure 11. The teal-colored violin (  ) and the orange-colored violin (  ) show the code coverage achieved by the original tests ($UT_0$) and the Baseline, respectively. The remaining four violins represent the UTFɪx-based techniques: UTFɪx$_{nc}$, UTFɪx$_s$, UTFɪx$_d$, and UTFIx$_{s+d}$ shown in 🟣 , 🟣 , 🟢 , and 🟡 violins, respectively. These techniques reach high code coverage percentages, with many tests clustering near 100% code coverage.

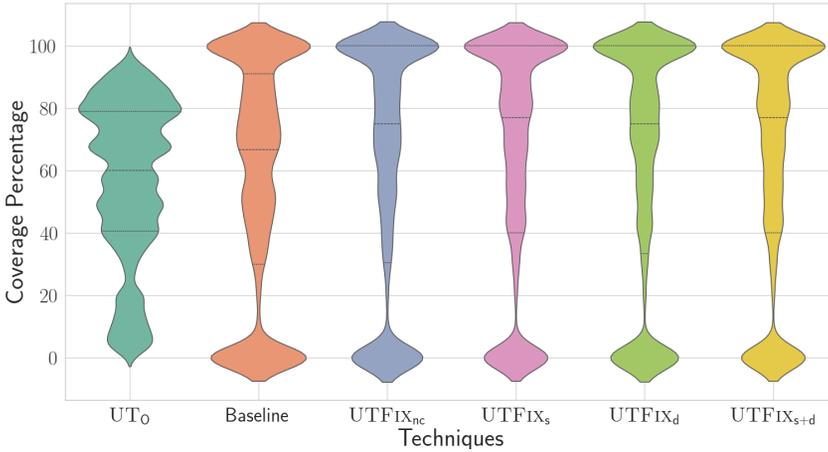

Fig. 11. Coverage results compared with the original test ($UT_0$) and the Baseline. It is noteworthy that the Baseline is $FM_1$ + error + $UT_0$, as defined in Section 5.4.

From the figure, we see that the median code coverage percentage by the $UT_0$, and Baseline are 60.00%, and 66.67%, respectively, showing the middle dashed line in the violin. Additionally, out of 369 tests, $UT_0$ does not achieve 100% code coverage in focal method with any test, with an average code coverage of 56.44%. In contrast, the median values for UTFIx$_{nc}$, UTFIx$_s$, UTFIx$_d$, and UTFIx$_{s+d}$ are 75.00%, 76.92%, 75.04%, and 76.83%, respectively. UTFIx$_s$ is able to achieve 100% code coverage for 96 tests.

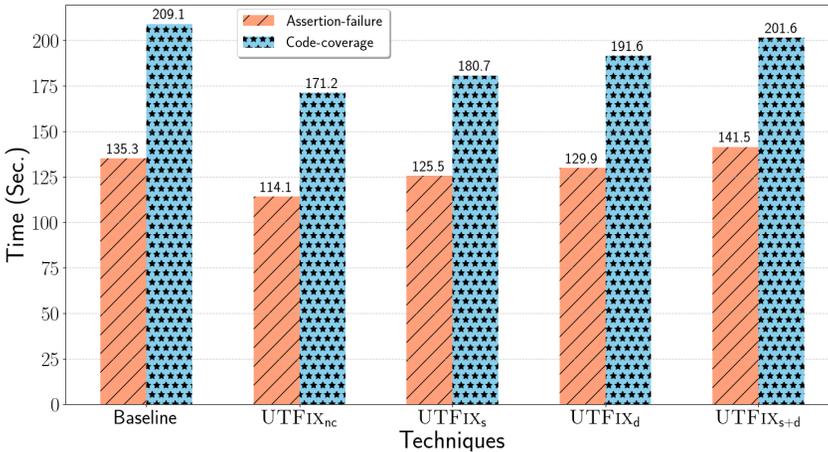

Fig. 12. Runtime of repaired tests for assertion failures (orange color with a double slash pattern (//)), and reduced code coverage (blue color with a star pattern (*)) caused by focal method changes.

## 6.3 RQ3: Runtime

Figure 12 shows the average runtime of repairing tests when having assertion failure and reduced code coverage due to the changes in the focal methods. The X-axis represents the four different settings of UTFIx, along with the Baseline. The Y-axis represents the time required to repair the tests, measured in seconds. The bar with a double slash pattern (//) represents the results for assertion failure, while the bar with a star pattern (*) represents the results for code coverage.

The average runtime of Baseline, $UTFix_{nc}$, $UTFix_s$, $UTFix_d$, and $UTFix_{s+d}$ for repairing tests with assertion failures is 135.3, 114.1, 125.5, 129.9, and 141.5 seconds, respectively. On the other hand, Baseline, $UTFix_{nc}$, $UTFix_s$, $UTFix_d$, and $UTFix_{s+d}$ take 209.1, 171.2, 180.7, 191.6, and 201.6 seconds, respectively, to improve the code coverage. The runtime for repairing tests to improve code coverage is higher compared to test repair during assertion failure. This is because achieving higher code coverage generally requires invoking a higher COT count which is set to 10. As shown in Figure 14b, many tests reached the maximum COT count to maximize code coverage. In contrast, the maximum count of COT for repairing tests that fail due to assertion failures is 5, as shown in Figure 14a. For the same reason, we also observe that the runtime of the Baseline is higher compared to the maximum time required by UTFix with dynamic slices. This is because, despite not using any static slices or dynamic slices, the Baseline fails to repair many tests. Consequently, for these tests, the Baseline continues until it reaches the maximum COT count.

## 6.4 RQ4: Diversity of Changes in Benchmark Creation and Test Repairs

Our benchmark dataset encompasses 51 different types of focal method changes—including API modifications, error-handling adjustments, and more—that results is assertion failures and reduced code coverage. This diversity ensures that the dataset reflects various real-world challenges rather than being tailored to a specific approach. For example, one such change involves error-handling in Figure 13a(i), where the focal method adds a check for empty enterprise account IDs, raising a ValueError at Line 4. UTFix does not rely on any assumptions about the type of changes; it simply focuses on the location of changes and the nature of the resulting errors to repair unit tests. When repairing the tests, we find the major changes in unit tests include changing assertions, changing expected values, adjusting assertion logic, adding or deleting assertions, handling exceptions, and incorporating mock objects.

Figure 13b and Figure 13c show how our tests are repaired during the assertion failure. The code on the left represents the changed focal methods, while the code on the right represents the corresponding repaired tests. In the changed focal method of Figure 13b(i), the same method "api.delete()" is invoked a second time at Line 3. However, the original unit test was only verifying that the method was invoked once, as indicated in Line 5 of Figure 13b(ii), which results in an assertion failure. To resolve this issue, UTFix automatically updates the assertion at Line 6 in Figure 13b(ii), ensuring the test passes by accurately validating the changed focal method's behavior. Similarly, Figure 13c(ii) shows the test repair by updating the expected values. Figure 13c(i) shows the changed focal method, which introduces a branch such that if the "account_id" is 'entUB', it returns an 'invalid_id' at Line 4. However, the test code expects 'entUB', resulting in an assertion failure at Line 7 in Figure 13c. UTFix automatically repairs the test by modifying the expected value for this assertion at Line 8.

Figure 13d shows the changes in the focal method and the corresponding updates in the test to improve code coverage. In the changed focal method, a new branch is introduced at Line 4. To cover this branch, UTFix proposes changes at Lines 8 and 10 in the test code. The focal method checks whether the number of arguments is two or more; therefore, UTFix introduces multiple assertions with both two and three arguments, ensuring full code coverage of the focal method.

We find one test may require one or more types of changes to be repaired. The most frequent change is 'changing assertions', accounting for 50.2%. Following this, 'expected value changes' account for 26.01% of cases, while 'adding new assertions' accounts for 9.3%.

```
1 def enterprise(self,
2    account_id: str) -> Enterprise:
3 +  if not account_id:
4 +    raise ValueError("NA")
5 return Enterprise(self,
6        account_id)
```

(i) Changed focal method

```
1 def test_enterprise(api):
2 +  with pytest.raises(ValueError,
3 +    match="NA")
4 +    api.enterprise("")
5 valid_id = "en"
6 enterprise = api.enterprise(valid_id)
7 assert enterprise.id == "en"
```

(ii) Repaired Test

(a) Example of Value Error

```
1 def delete(self) -> None:
2   self.api.delete(self.meta_url())
3 + self.api.delete(self.meta_url())
```

(i) Changed focal method

```
1 def test_delete(base, requests_mock):
2   m = requests_mock.delete(base.
3   meta_url(), json={..})
4   base.delete()
5 - assert m.call_count == 1
6 + assert m.call_count == 2
```

(ii) Repaired Test

(b) Example of Assertion Modification

```
1 def enterprise(self, account_id:
2   str) -> Enterprise:
3 + if account_id == "entUB":
4 +   return Enterprise(self,
5 +   "invalid_id")
6 return Enterprise(self,
7   account_id)
```

(i) Changed focal method

```
1 def test_enterprise(api, requests_mock,
2   sample_json):
3   url = api.build_url("meta/entUB")
4   requests_mock.get(url, json =
5   sample_json("EnterpriseInfo"))
6   enterprise = api.enterprise("entUB")
7 - assert enterprise.id == "entUB"
8 + assert enterprise.id == "invalid_id"
```

(ii) Repaired Test

(c) Example of Expected Values

```
1 def build_url(self, *component: str):
2 - return posixpath.join(*component)
3 + url = posixpath.join(*component)
4 + if len(component) > 2:
5 +   url +=f"?extra_param =
6 +   {len(component)}"
7 + return url.rstrip('/')
```

(i) Changed focal method

```
1 def test_build_url():
2   api = Api('apikey', 'https://../')
3 + rv = api.build_url('app', 'tbl')
4   assert rv == 'https://../v0/
5     app/tbl'
6 # Test with 3 components
7 + rv = api.build_url('app', 'tbl',
8 +   'rec')
9 + assert rv == 'https://../rec?
10 +   extra_param = 3'
```

(ii) Repaired Test

(d) Example of Branch Coverage

Fig. 13. Comparison of different changes in focal method and the corresponding repair

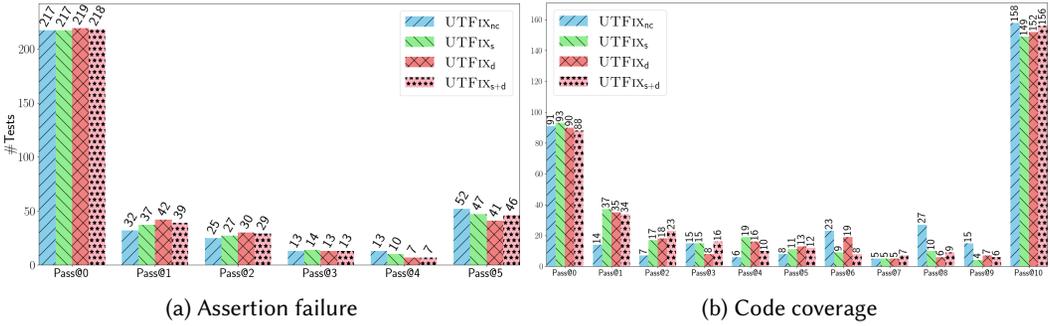

(a) Assertion failure  (b) Code coverage

Fig. 14. Comparison of COT distribution to repair tests that suffer from assertion failures and reduced code coverage. The horizontal axis represents the number of COT needed to repair the tests. The vertical axis represents (a) the number of tests repaired by UTFɪx, (b) last iteration that led to an improvement in code coverage by UTFɪx. Figure (a) shows that the maximum number of tests are repaired for assertion failures at Pass@0. Figure (b) shows that a large number of tests run till COT= 10 to improve the code coverage. It is noteworthy that tests reaching the maximum threshold, such as at Pass@5 and Pass@10, may or may not undergo test repair or show any improvement in code coverage. However, we include them to show how many tests reach the COT threshold.

## 6.5 RQ5: Effectiveness of UTFɪx in Repairing Real-World Change Aware Unit Tests

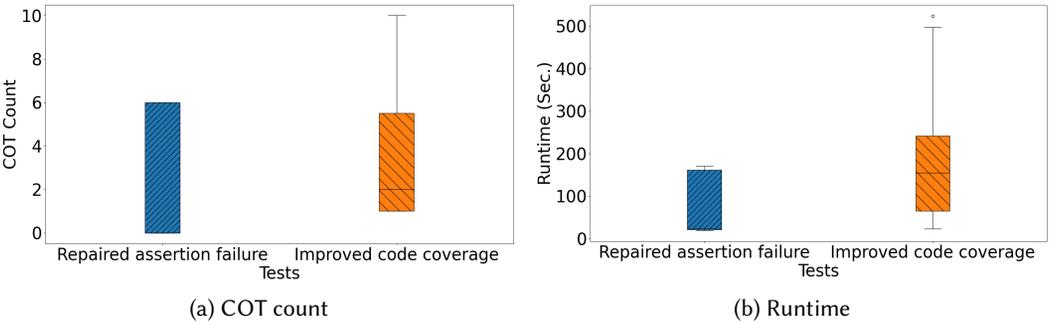

(a) COT count  (b) Runtime

Fig. 15. Resource usage in terms of feedback count (COT) and time required to repair the unit test for real-world data.

Our results show that UTFɪx is able to repair 12 tests that have assertion failures and improves code coverage for 19 tests. This highlights UTFɪx's ability to automatically repair tests. To further evaluate UTFɪx's efficiency, we measure both the chain-of-thought (COT) count—the number of feedback iterations—and the runtime, for both repairing assertion failures and improving code coverage.

Figure 15a compares the COT count between the two scenarios. The Figure shows that the median COT count is zero to repair tests that suffer from assertion failures, indicating that UTFɪx frequently resolves the failure in the first attempt without requiring additional feedback iterations, highlighting its efficiency in handling simpler test repairs. In the case of improving code coverage, the median COT count is four, showcasing UTFɪx's ability to improve code coverage with only a few feedback iterations, with a maximum COT count of 10.

Figure 15b shows the runtime required to repair assertion failures and improve code coverage. The median runtime to repair tests that suffer from assertion failures is 23.08 seconds. In contrast, the median runtime to improve code coverage is approximately 160 seconds, with some cases reaching up to 500 seconds.

## 7 Discussion

**Assumption of Focal Method Correctness:** In real-world software evolution, both the focal method and the associated unit test can be incorrect. However, UTFix operates in a regression setting, assuming that focal method has evolved correctly. Our evaluation data is curated based on this assumption, and the experiments are conducted accordingly. Thus, ideally, UTFix should repair all the assertion failures as such failures usually reflect the actual behavior and the expected value differ, highlighting potential bugs or unsatisfied requirements. Since we are assuming the focal method is correct, the bugs are thus coming from the unit test.

**Preventing Data Leakage:** UTFix is primarily evaluated on uniquely changed focal methods generated in-house, ensuring they are new, distinct, and not publicly available. Since our subject projects are collected from open-source GitHub repositories, we also validate the uniqueness of the changed focal methods by randomly sampling 20 examples from five projects and thoroughly reviewing their commit histories using keyword searches and manual verification. No exact matches of the focal method diffs were found in prior commits, meaning that the specific code changes used in our dataset have not appeared before. This minimizes the risk of overlap with any data seen during the training of LLMs like Claude 3.5 Sonnet.

**Ensuring Repair Reliability:** In UTFix, a 'false positive' repair might be considered if a test is updated and marked as repaired without truly repairing the error. To ensure repair reliability, we run each test 10 times, confirming consistent passing results. Additionally, we manually review a sample of 10 repairs to verify their effectiveness and confirm no 'false positives', ensuring UTFix genuinely addresses the issues in the original tests.

**Use of Open-Source LLM:** We conducted experiments using open-source models (Llama and DeepSeek-Coder) from Huggingface [2] following the same experimental setup as UTFix. This includes the same structured prompts with the same content across four prompt settings -no slice (UTFix$_{nc}$), static slices (UTFix$_s$), dynamic slices (UTFix$_d$), and a combination of static and dynamic slices (UTFix$_{s+d}$)- as well as COT using five times feedback.

For the Llama evaluation, we use the latest version of the Llama family models, specifically the Llama3-8b. Our findings indicate that Llama3-8b with UTFix$_d$ achieves the highest repair rate for assertion failures among all prompt settings, 38.1% (134/352) of tests, compared to a baseline (i.e., Llama3-8b with the changed focal method) repair rate of 27.3% (96/352). For improving code coverage, Llama3-8b with UTFix$_s$ is the most effective, achieving 100% code coverage in 46 tests, compared to the baseline of 34 tests.

We conducted experiments with DeepSeek-Coder and found that deepseek-ai/DeepSeek-Coder-V2-Instruct requires substantial GPU resources, needing at least 8 GPUs with 80GiB each for the generation task as it is a 236B MoE model. Due to this limitation, we use deepseek-ai/DeepSeek-Coder-V2-Lite-Instruct, which is less resource-intensive and has 16B total parameters with 2.4B active parameters. Using deepseek-ai/DeepSeek-Coder-V2-Lite-Instruct with UTFix$_d$, we achieve the highest repair rate for assertion failures among all prompt settings, 22.4% (79/352), compared to a baseline repair rate of 15.1% (53/352). Furthermore, to improve code coverage, deepseek-ai/DeepSeek-Coder-V2-Lite-Instruct with UTFix$_s$ is the most effective, achieving 100% code coverage in 36 tests, compared to the baseline of 21 tests. More detailed results are available in [34].

Although the model performance is lower—mostly due to the smaller size of these open-source models—the results show a similar trend to our previously reported results, i.e., our main findings generalize to open-source LLMs. Specifically, context collection using dynamic and static slicing improves LLMs' effectiveness in test repair.

## 8 Threats to Validity

*External Validity.* Our study focuses on open-source Python projects, which may limit the generalizability of the results to other programming languages and environments. While Python is widely used, the effectiveness of UTFix on projects written in languages with different testing frameworks, paradigms, or architectures (e.g., statically typed languages like Java or C++) may differ. However, the core principles and methodology of UTFix are language-agnostic, as they focus on analyzing and repairing tests along with their corresponding focal methods. So, with appropriate datasets and adaptations for language-specific syntax and testing frameworks, UTFix can be extended to support other programming languages.

Our real-world benchmark dataset is derived from a set of open-source projects with specific characteristics (e.g., popular, actively maintained). This introduces a potential sampling bias, as smaller or less active projects may face different challenges that were not captured in our evaluation. The benchmark datasets may not represent all possible code evolution patterns, limiting the comprehensiveness of our results.

*Internal Validity.* The synthetic benchmark dataset that we curated involves injecting realistic changes into focal methods and evaluating whether UTFix can repair the unit tests. However, the artificial nature of these changes may not perfectly replicate real-world software evolution. Although we also evaluated UTFix on real-world changes, the number of sampled real changes may not capture all possible scenarios.

Furthermore, the assumption that naming conventions (e.g., similarity between test name and focal methods) can always help establish <FM, UT> pairs may not hold for all projects. To address potential concerns that name similarity could introduce bias in the repair process, we conducted an additional experiment where we replaced the test name in a unit test using a combination of random string generation and regex-based substitution. We then applied UTFix to determine whether the unit test could still be repaired. Our results indicate that UTFix successfully repaired 99.34% of the tests compared to the original repair rate and achieved 100% code coverage for 96 tests, the same as the original repair.

UTFix restricts test repair to within the body of the test code. So, UTFix fails to repair tests if the repair falls outside the test code. This may happen for parameterized tests where expected values are defined externally or tests are reliant on external setups, such as mock objects. Expanding UTFix's repair scope to handle these external setups could address such issues. Furthermore, certain repairs may fall outside the scope of automated solutions, particularly when the specifications of the underlying focal method are unknown or cannot be inferred. In such cases, involving a human in the loop will be essential.

*Construct Validity.* The way we measure the effectiveness of UTFix—through assertion failure and code coverage—could be limiting. Although these metrics are widely used in the software testing literature, they may not capture all aspects of test quality, such as readability, maintainability, or the comprehensiveness of the repaired tests. While other metrics such as mutation scores could provide additional insights, we believe that concentrating on the primary factors offers a more accurate assessment of UTFix's performance in a differential testing context. This approach ensures that our evaluation aligns with the core objective of verifying UTFix's capability to address immediate test failures resulting from code changes.

UTFix relies on LLMs for generating test repairs. The quality and reliability of these repairs depend heavily on the model's training data and its ability to interpret the context correctly. LLMs sometimes tend to produce identical outputs for the same input. We address this by carefully tuning hyperparameters such as temperature, which enhances diversity in the generated responses. Additionally, we structure our feedback to the model in a way that clarifies the intended output, ensuring that if an initial response is incorrect, the model can adjust and generate more accurate results in subsequent attempts. Further, we rely on a single LLM due to cost constraints. In the future, we plan to reproduce the results with different LLMs.

## 9 Related Work

Repairing broken unit tests is an important task as raised in [27], [11], [42], [30], [7]. However, despite its importance, there are limited studies in the literature that systematically benchmark and address these issues.

Daniel et al. proposed ReAssert [11], a tool for automatically repairing broken JUnit tests by applying heuristic-based repair techniques, such as altering expected values in assertion statements. Later, they extended ReAssert by incorporating symbolic execution for test repair [10]. In this approach, the expected value in an assertion is modified based on solutions to symbolic constraints derived from literals affecting the expected value. This ensures that the actual computed value remains unchanged. If a solution to the symbolic constraints is found, the tool updates the literal values in the unit test. Otherwise, the test is deemed irreparable.

Mirzaaghaei et al. [24] introduced TestCareAssistant (TCA), a framework designed to either repair broken tests or generate new ones based on existing tests. This framework primarily targets a limited set of changes, such as additions of method parameters. The repair process modifies variables and values within the test to explore possible fixes. Xu et al. developed TestFix, which employs a genetic algorithm to repair failing unit tests [40]. This approach identifies the optimal sequence of method call insertions and deletions required to restore a functional test. However, TestFix is limited in scope, as it only repairs tests containing a single assertion statement. To address this limitation, Li et al. introduced TRIP, a technique that prioritizes preserving the original intent of a test while performing test repair [21]. TRIP leverages a search-based approach to generate repair candidates by updating the unit test and ranking them based on dynamic symbolic execution.

Given the recent impressive performance of Large Language Models (LLMs) [28, 29, 43], researchers have started exploring their potential for test repair [19, 23, 31, 32, 37] using LLMs. Yaraghi et al. introduced TARGET for JUnit test repair [42]. TARGET analyzes changes in the system under test (SUT) and detects test breakage locations. It prioritizes repair context using call graph analysis, refining the context further using TF-IDF similarity between changed lines in the SUT and broken test lines. After prioritization, a fine-tuned pre-trained CodeT5 model generates test repairs efficiently.

Due to the widespread adoption of LLMs, researchers have also investigated automated test generation through both search-based software testing and LLM-based approaches [13–15, 19, 23, 26, 32, 37, 39]. SymPrompt, for example, uses path constraints from symbolic execution to generate effective prompts for LLMs, capturing execution paths to generate effective unit tests [31]. It considers argument types, external dependencies, and method behavior to refine generated tests iteratively. Schafer et al. proposed TESTPILOT, an adaptive LLM-based test generation tool for JavaScript unit tests [32]. TESTPILOT employs API exploration, documentation mining, and iterative prompting strategies to improve test accuracy.

Another test generation technique, TELPA, was proposed by Yang et al. [41]. TELPA focuses on generating tests for hard-to-cover branches using a combination of program analysis and counter-example sampling. It collects uncovered branches and reuses existing tests as counter-examples to achieve greater coverage. Similarly, Alshahwan et al. introduced TestGenLLM, an approach designed to improve the reliability and effectiveness of LLM-generated tests by emphasizing test validation and filtering [7]. Their method aims to eliminate flaky tests and improve code coverage for Meta.

TestART generates unit tests for a given focal method, considering them as valid if they compile, although its experiments are limited to only five projects [18]. Unlike TestART, which creates new tests from scratch, our proposed approach repairs and adapts existing tests to reflect SUT changes more accurately. Similarly, TOGA generates test oracles and assertions to detect bugs in focal method [12]. However, TOGA focuses only on assertions, relying on external tools like EvoSuite for the rest of the test structure. In contrast, our approach repairs tests holistically by modifying both assertions and test setup, thereby achieving a higher repair rate.

## 10 Conclusions

In this paper, we propose UTFix, a novel approach designed to automatically repair unit tests when the corresponding focal method undergoes changes. By leveraging LLMs, UTFix tackles two critical challenges: assertion failure and reduced code coverage, both of which can arise from code changes. UTFix integrates both static and dynamic information to guide LLMs in generating meaningful and syntactically correct repairs, offering an automated solution to maintain the reliability of evolving software.

We evaluated UTFix on both synthetic and real-world benchmark datasets. Our synthetic dataset demonstrates UTFix's ability to repair up to 89.2% of assertion failures and significantly improve code coverage, while real-world examples also show promising results in addressing assertion failures and coverage reduction in actual change aware settings. These evaluations highlight the potential of UTFix in practical, large-scale software development environments, where keeping tests up to date is critical but often neglected.

Our contributions include the development of an automated repair technique, the curation of comprehensive benchmark datasets, and the implementation of UTFix for Python projects. Future work will explore extending this approach to more programming languages and refining the use of LLMs to handle more complex test repair scenarios.

## Data-Availability Statement

Our artifact is available in [34] that includes source code, benchmark datasets, and repaired tests.

## References

[1] 2024. Claude-3.5-sonnet. https://www.anthropic.com/news/claude-3-5-sonnet.

[2] 2024. huggingface. https://huggingface.co/.

[3] 2024. langchain. https://api.python.langchain.com/en/latest/chat_message_histories/langchain_community.chat_message_histories.in_memory.ChatMessageHistory.html.

[4] 2024. Tox. https://tox.wiki/en/latest/config.html.

[5] Mithun Acharya and Brian Robinson. 2011. Practical change impact analysis based on static program slicing for industrial software systems. In *International Conference on Software Engineering*. 746–755.

[6] Hiralal Agrawal and Joseph R Horgan. 1990. Dynamic program slicing. *ACM SIGPlan Notices* 25, 6 (1990), 246–256.

[7] Nadia Alshahwan, Jubin Chheda, Anastasia Finogenova, Beliz Gokkaya, Mark Harman, Inna Harper, Alexandru Marginean, Shubho Sengupta, and Eddy Wang. 2024. Automated unit test improvement using large language models at meta. In *International Symposium on Foundations of Software Engineering*. 185–196.

[8] Tatiana Avdeenko and Konstantin Serdyukov. 2021. Automated test data generation based on a genetic algorithm with maximum code coverage and population diversity. *Applied Sciences* 11, 10 (2021), 4673.

[9] M-H Chen, Michael R Lyu, and W Eric Wong. 2001. Effect of code coverage on software reliability measurement. *IEEE Transactions on reliability* 50, 2 (2001), 165–170.

[10] Brett Daniel, Danny Dig, Tihomir Gvero, Vilas Jagannath, Johnston Jiaa, Damion Mitchell, Jurand Nogiec, Shin Hwei Tan, and Darko Marinov. 2011. Reassert: a tool for repairing broken unit tests. In *International Conference on Software Engineering (Tool Demonstrations Track)*. 1010–1012.

[11] Brett Daniel, Vilas Jagannath, Danny Dig, and Darko Marinov. 2009. ReAssert: Suggesting repairs for broken unit tests. In *International Conference on Automated Software Engineering*. 433–444.

[12] Elizabeth Dinella, Gabriel Ryan, Todd Mytkowicz, and Shuvendu K Lahiri. 2022. Toga: A neural method for test oracle generation. In *International Conference on Software Engineering*. 2130–2141.

[13] Gordon Fraser and Andrea Arcuri. 2011. Evosuite: automatic test suite generation for object-oriented software. In *International Symposium on Foundations of Software Engineering*. 416–419.

[14] Gordon Fraser and Andrea Arcuri. 2013. Evosuite: On the challenges of test case generation in the real world. In *International Conference on Software Testing, Verification, and Validation*. 362–369.

[15] Gordon Fraser and Andrea Arcuri. 2014. A large-scale evaluation of automated unit test generation using evosuite. *ACM Transactions on Software Engineering Methodology* 24, 2 (2014), 1–42.

[16] Malcom Gethers, Bogdan Dit, Huzefa Kagdi, and Denys Poshyvanyk. 2012. Integrated impact analysis for managing software changes. In *International Conference on Software Engineering*. 430–440.

[17] Rahul Gopinath, Carlos Jensen, and Alex Groce. 2014. Code coverage for suite evaluation by developers. In *International Conference on Software Engineering*. 72–82.

[18] Siqi Gu, Chunrong Fang, Quanjun Zhang, Fangyuan Tian, Jianyi Zhou, and Zhenyu Chen. 2024. Improving LLM-based Unit test generation via Template-based Repair. *arXiv preprint arXiv:2408.03095* (2024).

[19] Sepehr Hashtroudi, Jiho Shin, Hadi Hemmati, and Song Wang. 2023. Automated test case generation using code models and domain adaptation. *arXiv preprint arXiv:2308.08033* (2023).

[20] Mansi Khemka and Brian Houck. 2024. Toward Effective AI Support for Developers: A survey of desires and concerns. *Commun. ACM* 67, 11 (2024), 42–49.

[21] Xiangyu Li, Marcelo d'Amorim, and Alessandro Orso. 2019. Intent-preserving test repair. In *International Conference on Software Testing, Verification, and Validation*. 217–227.

[22] Yue Liu, Thanh Le-Cong, Ratnadira Widyasari, Chakkrit Tantithamthavorn, Li Li, Xuan-Bach D Le, and David Lo. 2024. Refining chatgpt-generated code: Characterizing and mitigating code quality issues. *ACM Transactions on Software Engineering and Methodology* 33, 5 (2024), 1–26.

[23] Stephan Lukasczyk and Gordon Fraser. 2022. Pynguin: Automated unit test generation for python. In *International Conference on Software Engineering Companion*. 168–172.

[24] Mehdi Mirzaaghaei. 2011. Automatic test suite evolution. In *International Symposium on Foundations of Software Engineering*. 396–399.

[25] Mehdi Mirzaaghaei, Fabrizio Pastore, and Mauro Pezzè. 2012. Supporting test suite evolution through test case adaptation. In *International Conference on Software Testing, Verification, and Validation*. 231–240.

[26] Carlos Pacheco, Shuvendu K Lahiri, Michael D Ernst, and Thomas Ball. 2007. Feedback-directed random test generation. In *International Conference on Software Engineering*. 75–84.

[27] Strategic Planning. 2002. The economic impacts of inadequate infrastructure for software testing. *National Institute of Standards and Technology* 1 (2002).

[28] Shanto Rahman, Abdelrahman Baz, Sasa Misailovic, and August Shi. 2024. Quantizing large-language models for predicting flaky tests. In *International Conference on Software Testing, Verification, and Validation*. 93–104.

[29] Shanto Rahman, Bala Naren Chanumolu, Suzzana Rafi, August Shi, and Wing Lam. 2025. Ranking Relevant Tests for Order-Dependent Flaky Tests. In *International Conference on Software Engineering*.

[30] Shanto Rahman and August Shi. 2024. FlakeSync: Automatically Repairing Async Flaky Tests. In *International Conference on Software Engineering*. 1–12.

[31] Gabriel Ryan, Siddhartha Jain, Mingyue Shang, Shiqi Wang, Xiaofei Ma, Murali Krishna Ramanathan, and Baishakhi Ray. 2024. Code-Aware Prompting: A Study of Coverage-Guided Test Generation in Regression Setting using LLM. In *International Symposium on Foundations of Software Engineering*. 416–971.

[32] Max Schäfer, Sarah Nadi, Aryaz Eghbali, and Frank Tip. 2023. An empirical evaluation of using large language models for automated unit test generation. *IEEE Transactions on Software Engineering* (2023).

[33] Dávid Tengeri, Árpád Beszédes, Tamás Gergely, László Vidács, Dávid Havas, and Tibor Gyimóthy. 2015. Beyond code coverage—An approach for test suite assessment and improvement. In *International Conference on Software Testing, Verification and Validation Workshops*. 1–7.

[34] UTFix 2025. https://sites.google.com/view/utfix.

[35] Arash Vahabzadeh, Amin Milani Fard, and Ali Mesbah. 2015. An empirical study of bugs in test code. In *International Conference on Software Maintenance and Evolution*. 101–110.

[36] Zejun Wang, Kaibo Liu, Ge Li, and Zhi Jin. 2024. HITS: High-coverage LLM-based Unit Test Generation via Method Slicing. In *International Conference on Automated Software Engineering*. 1258–1268.

[37] Jason Wei, Xuezhi Wang, Dale Schuurmans, Maarten Bosma, Fei Xia, Ed Chi, Quoc V Le, Denny Zhou, et al. 2022. Chain-of-thought prompting elicits reasoning in large language models. *Advances in neural information processing systems* 35 (2022), 24824–24837.

[38] Mark Weiser. 1984. Program slicing. *IEEE Transactions on software engineering* 4 (1984), 352–357.

[39] Zhuokui Xie, Yinghao Chen, Chen Zhi, Shuiguang Deng, and Jianwei Yin. 2023. ChatUniTest: a ChatGPT-based automated unit test generation tool. *arXiv preprint arXiv:2305.04764* (2023).

[40] Yong Xu, Bo Huang, Guoqing Wu, and Mengting Yuan. 2014. Using genetic algorithms to repair JUnit test cases. In *Asia-Pacific Software Engineering Conference*. 287–294.

[41] Chen Yang, Junjie Chen, Bin Lin, Jianyi Zhou, and Ziqi Wang. 2024. Enhancing LLM-based Test Generation for Hard-to-Cover Branches via Program Analysis. *arXiv preprint arXiv:2404.04966* (2024).

[42] Ahmadreza Saboor Yaraghi, Darren Holden, Nafiseh Kahani, and Lionel Briand. 2024. Automated Test Case Repair Using Language Models. *IEEE Transactions on Software Engineering* (2024).

[43] Jialu Zhang, Todd Mytkowicz, Mike Kaufman, Ruzica Piskac, and Shuvendu K Lahiri. 2022. Using pre-trained language models to resolve textual and semantic merge conflicts (experience paper). In *International Symposium on Software Testing and Analysis*. 77–88.